%% file: edsos2014_tp.tex
\documentclass[conference]{IEEEtran}

\usepackage[draft, index, inline, nomargin]{fixme}
\fxuseenvlayout{plain}
\fxusetheme{color}

\usepackage{graphicx}
\usepackage[table]{xcolor}
\usepackage{color}
\usepackage{wrapfig}
\usepackage{pgf}
\usepackage{float}
\usepackage{caption}
\usepackage{subcaption}
\usepackage{xspace}

\usepackage{listings}
\usepackage{cml_def}
\lstdefinestyle{cmlLanguageStyle}{basicstyle=\footnotesize\ttfamily,
			frame=trBL, 
			showstringspaces=false,  
			captionpos=b,
			frameround=tttt, 
			aboveskip=5mm, 
			belowskip=5mm,
			breaklines=true,
			tabsize=4,
			framexleftmargin=0mm, 
			framexrightmargin=0mm}
\lstnewenvironment{cml}[1][]{\lstset{style=cmlLanguageStyle}\lstset{language=CML}#1}
{}

\usepackage[colorlinks=true,linkcolor=black, bookmarksnumbered=true]{hyperref}

\usepackage{listings}  

\usepackage{acronym}

\acrodef{cml}[CML]{COMPASS Modelling Language}
\acrodef{pog}[POG]{Proof Obligation Generator}
\acrodef{po}[PO]{Proof Obligation}
\acrodef{pos}[POs]{Proof Obligations}
\acrodef{tpp}[TPP]{Theorem Prover Plugin}
\acrodef{ast}[AST]{Abstract Syntax Tree}
\acrodef{sos}[SoS]{System of Systems}
\acrodefplural{sos}{Systems of Systems}
\acrodef{utp}[UTP]{Unifying Theories of Programming}
\acrodef{vdm}[VDM]{Vienna Development Method}
\def\vdm{VDM}

\def\toolpl{Symphony tool platform}
\def\tool{Symphony}

\newcommand{\Circus}{\emph{Circus}\xspace}
\newcommand{\isabelleutp}{\emph{Isabelle/UTP}\xspace}

\title{Towards Verification of Constituent Systems \\ through Automated Proof}
\author{Lu\'{i}s Diogo Couto\inst{1} 
\and Simon Foster\inst{2}
\and Richard Payne\inst{3}}

\author{\IEEEauthorblockN{Lu\'{i}s Diogo Couto}
\IEEEauthorblockA{Aarhus University, Denmark\\
ldc@eng.au.dk} 
\and
\IEEEauthorblockN{Simon Foster}
\IEEEauthorblockA{University of York, United Kingdom\\
simon.foster@york.ac.uk}
\and
\IEEEauthorblockN{Richard Payne}
\IEEEauthorblockA{Newcastle University, United Kingdom\\
richard.payne@ncl.ac.uk}
}

\begin{document}
\maketitle

\begin{abstract} 
  This paper explores verification of constituent systems within the
  context of the \emph{Symphony} tool platform for Systems of Systems
  (SoS). Our SoS modelling language, CML, supports various contractual
  specification elements, such as state invariants and operation
  preconditions, which can be used to specify contractual obligations
  on the constituent systems of a SoS. To support verification of these obligations
  we have developed a proof obligation generator and theorem prover
  plugin for Symphony. The latter uses the \emph{Isabelle/HOL} theorem
  prover to automatically discharge the proof obligations arising from
  a CML model. Our hope is that the resulting proofs can then be used
  to formally verify the conformance of each constituent system,
  which is turn would result in a dependable SoS.
\end{abstract}

%resets the acronyms
\acresetall

\input{sections/introduction}

\input{sections/background}

\input{sections/related}

\input{sections/tppog}

\input{sections/example}

% \input{sections/discussion}

\input{sections/conclusion}

\section*{Acknowledgements}

This work is supported by EU Framework 7 Integrated Project
``Comprehensive Modelling for Advanced Systems of Systems'' (COMPASS, Grant
Agreement 287829). For more information see \url{http://www.compass-research.eu}.

\bibliographystyle{IEEEtran}
\bibliography{edsos2014_tp}

\end{document}

%% file: sections/introduction.tex
\section{Introduction}
%\fxnote{usual bit about compass and cml}\\

A System of Systems (SoS) \cite{Kopetz13} is a collection of
semantically heterogeneous, independent, and distributed constituent
systems (CSs) which are co-ordinated to achieve an overall
goal. Independence means that no CS can exert control on another CS,
only influence its behaviour by offering potential opportunities
should synergy be reached. Since CSs are dynamic and heterogeneous,
often changing their capabilities and services, such synergy is
achieved by negotiation of \emph{contracts} between a set of CSs,
which impose binding conditions on the behaviour of each CS. Since
failure of such an agreement will result in degradation of the SoS, it
is important that each CS has some measure of certainty in its ability
to fulfil its requirements, which in turn will lead to a dependable
SoS.

System of Systems Engineering (SoSE) therefore requires languages with
which we can accurately model CSs to predict their behaviour, and
tools which enable their verification. Such languages should have a
sound theoretical background to ensure that they can be assigned a
consistent behaviour, and the ability to handle the composition of
heterogeneous constituents. To this end the
\ac{cml}~\cite{Woodcock&12a} has been developed, a formal modelling
language for SoSs. \ac{cml} reproduces the style of the
VDM-SL~\cite{Jones90a} formal specification language, whilst
integrating CSP~\cite{Hoare85} process modelling constructs from the
\Circus~\cite{Woodcock01} language. \ac{cml} has a formal semantics
based in Hoare and He's \ac{utp}~\cite{Hoare98}, in its denotational,
operational, and axiomatic flavours. Along with the associated
\emph{Symphony}\footnote{Symphony can be downloaded from
  \url{http://symphonytool.org/}} tool platform, \ac{cml} allows SoSs
and CSs to be formally modelled, tested, and verified in a controlled
environment.

This paper focuses on two closely related components of Symphony, the
\ac{tpp} and \ac{pog}. A theorem prover can be applied to
\emph{verify} a software system, that is mathematically demonstrate
that required properties are met through mechanically verified
proof. In the case of CSs, we need to verify that the internal
functionality and pattern of interaction is guaranteed to fulfil the
contract. In Symphony this verification can be facilitated through the
\ac{pog} which generates proof goals upon which the correctness of the
CS model depends. \ac{cml} has a number of facilities for specifying
contractual obligations, such as type invariants, pre- and
post-conditions for functions and operations, and system state
invariants. These can then variously be used to specify contractual
obligations for a CS model, and the application of the \ac{pog} in
concert with the \ac{tpp} can be used to verify that the system
satisfies those obligations. Our thesis is, therefore, that these
technologies provide a way forward for mechanically verifying that a
CS model fulfils its contractual obligations to the wider SoS.

In the remainder we outline our
contributions. Section~\ref{sec:background} gives more background to
our baseline technology. Section~\ref{sec:related} discusses related
work. Section~\ref{sec:pog_tp} discusses the combined \ac{pog} and
\ac{tpp} framework, and how it can be used to verify a CS
model. Section~\ref{sec:example} demonstrates an example CS, and how
we envisage verifying it for a wider SoS. Finally in
%Section~\ref{sec:furtherwork} we outline future work and in
Section~\ref{sec:conc} we conclude and outline future work.

%% file: sections/background.tex
\section{Background}
\label{sec:background}

\ac{cml} is a language for modelling constituent systems and their
composition in an SoS. Systems are modelled using \ac{cml}
\emph{processes}, which are stateful reactive entities that can be
executed concurrently, and exchange messages over \emph{channels} in
the style of the CSP process calculus~\cite{Hoare85}. A \ac{cml} model
consists of a collection of user defined \emph{types},
\emph{functions}, \emph{channels}, and \emph{processes}. A process, in
turn, consists of private \emph{state variables}, \emph{operations}
that act on these variables, and \emph{actions} that specify reactive
behaviour using operators from CSP. \ac{cml} processes can be parallel
composed to represent concurrent execution, enabling description of a
complete SoS.

\ac{cml} has a formal mathematical foundation~\cite{Fitzgerald&13d}
based in the \ac{utp} semantic framework~\cite{Hoare98}, which allows
processes to be given a precise semantics. \ac{utp} allows us to
tackle semantic heterogeneity in SoSE by decomposing a modelling
language semantics into its theoretical building blocks, such as
state, concurrency, discrete time, and mobility, which can then be
formalised as \emph{``UTP theories''}. UTP theories then act as
components with which we can construct semantic models for languages
and provide links between similar languages based on common
theoretical factors.

\begin{figure}[t]
\begin{center}
  \includegraphics[width=8.8cm]{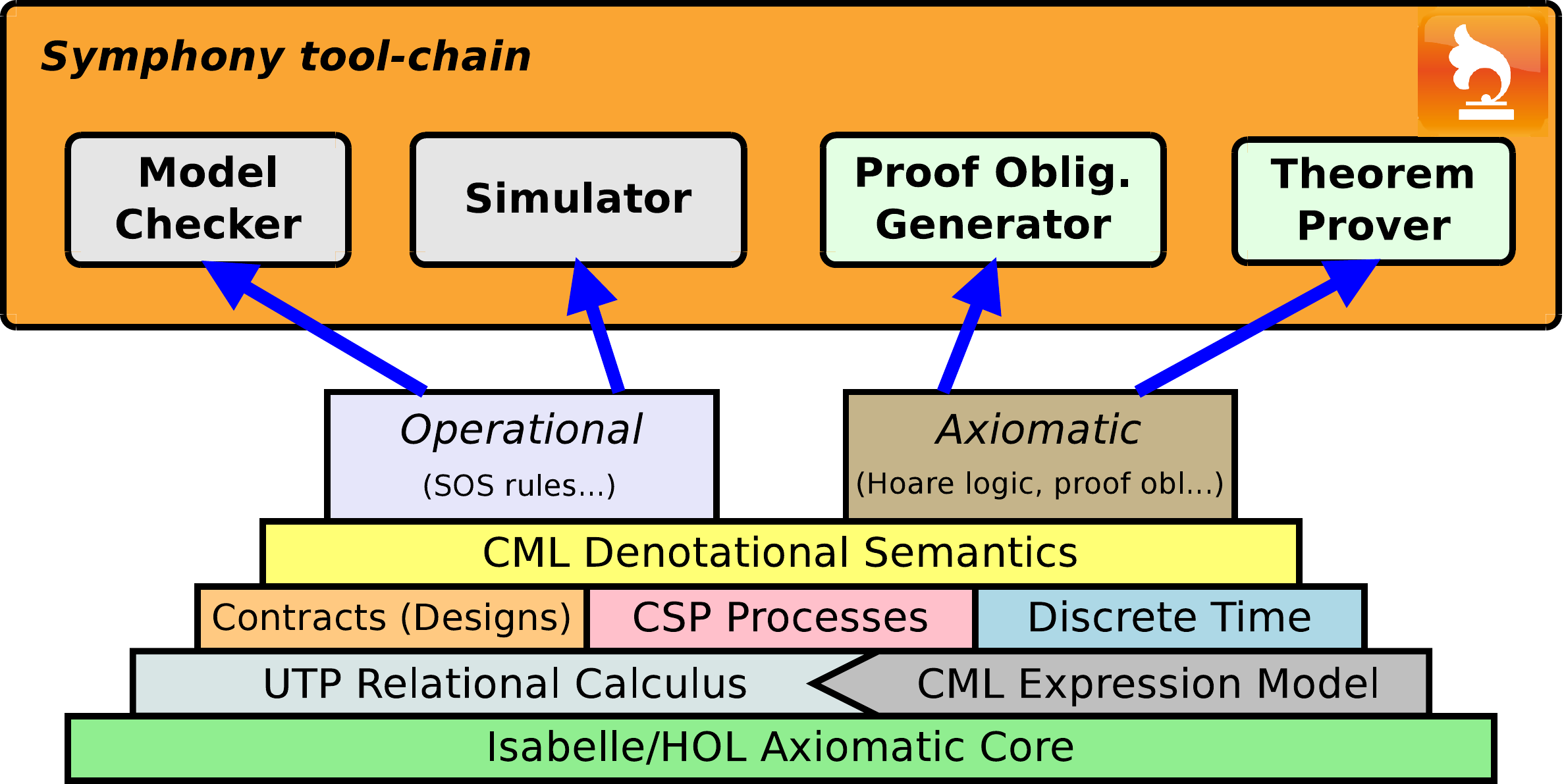}
  \caption{Semantically supported Symphony tool platform}
  \label{fig:cmlstack}
\end{center}
\vspace{-5.5ex}
\end{figure}

Development of \ac{cml} models is aided through the associated
\emph{Symphony} tool platform. Symphony is an Eclipse-based
development environment that provides a parser, syntax highlighting, a
type checker, simulator, model checker, and a variety of other tools
for variously constructing and verifying \ac{cml} models. Though
consisting of independently developed components, the different tools
share a common semantic foundation given by the UTP denotational
model. This ``semantic stack'' is shown in Figure~\ref{fig:cmlstack},
with the Symphony platform and associated components positioned on
top. Within the UTP relational calculus several computational
paradigms have been formalised as UTP theories (Contracts, Processes,
etc.), and these have been in turn composed to produce the CML
denotational semantics. Finally, reference semantics have been
produced that underlie the various tools, including the operational
semantics, which underlies the simulator and model checker, and
various axiomatic semantics, such as a Hoare
calculus~\cite{COMPASSD23.4d}. Since this formal link exists from each
tool down into the unified semantic basis we can have a degree of
certainty that the various evidences produced can be consistently
composed to verify a CS model.

To support such verifications we have created a theorem prover plugin,
based on the \emph{Isabelle/HOL}~\cite{Isabelle} interactive theorem
prover. Isabelle/HOL is ideal for this kind of verification since
proofs can be independently checked with respect to a secure axiomatic
core; a facet of the \emph{``LCF architecture''}. Our theorem prover
is based in a mechanised semantic framework for \ac{utp} called
\isabelleutp~\cite{Foster14} that provides a strong theoretical
grounding for CML, ensuring its consistency. We have mechanised a
partial semantic model for \ac{cml} in \isabelleutp, a collection of
associated proof tactics, and a visitor that translates the \ac{cml}
\ac{ast} into Isabelle definitions that can then be used to support
proof. Our current approach to proof in \ac{cml} is to, where
possible, convert \ac{cml} to equivalent HOL formulae, and perform the
proof using Isabelle's variety of existing tactics and laws,
effectively transferring results from HOL to UTP. In line with the UTP
framework, the theorem prover is fully \emph{extensible}: we can add
support for additional programming concepts, and associated tactics as
required in the future.

Alongside the theorem prover, Symphony contains a \ac{pog} that
generates, for a given CML document, a collection of proof goals that
must be satisfied to prove certain high level properties of the model,
such as internal consistency, contractual correctness of operations,
and termination. The \ac{tpp} can then be used to attempt discharge of
these proof obligations, resulting in concrete proof objects for the
properties. We are currently working towards a formal axiomatic
semantics for these proof obligations based on the current
implementation, which will allow the integration of these proof
objects with other evidences in the tool-chain.

% Since these proofs can be linked to the underlying
% semantic model in the \ac{utp}, they could be composed to construct a
% certificate showing conformance of the CS to the SoS contract.

%% file: sections/related.tex
\section{Related Work}
\label{sec:related}

The Symphony tool platform is an extension of the open source
\emph{Overture} IDE~\cite{Larsen&10a} for VDM based modelling. The
\tool\ \ac{pog} is an adaptation of the \ac{pog} for
Overture~\cite{couto&13a} to also handle CML proof
obligations. Previous efforts to generate and discharge \acp{po} for
\vdm\ include~\cite{Agerholm&99b} and~\cite{Vermolen07}, which connect
VDMTools and Overture \acp{po} respectively to the HOL4 theorem
prover~\cite{HOL4}. These attempts were limited to the functional
subset of VDM. We use a similar mapping for \ac{cml} types and
expressions, whilst adding support for \ac{cml}'s imperative and
concurrent constructs.

The area of theorem proving tools includes a number of options including
Isabelle/HOL~\cite{Isabelle} (which we use);
PVS\footnote{\url{http://pvs.cdl.sri.com}} combining a specification language
with a theorem prover; Coq~\cite{bertot2004interactive}, a proof
assistant based on intuitionistic logic; specialised verification systems such
as \emph{Spec$\sharp$}, which is based on the Boogie verification
language~\cite{Barnett&06} and supported by the \emph{Z3} SMT
solver~\cite{DeMoura08}; and the
\emph{Rodin}\footnote{\url{http://event-b.org}} tool for Event-B which includes
an automated theorem prover.

We choose Isabelle for several reasons. It is based on \emph{Higher
  Order Logic} which is ideal for embedding a language like
\ac{cml}. The LCF architecture ensures proofs are
correct with respect to a secure logical core. It has
a large library of mathematical structures related to program
verification, such as relational calculus and lattice theory. It
integrates powerful proof facilities, such as the \textsf{auto} tactic
for automated deduction, integration of first-order automated theorem
provers (like \emph{Z3}) in the \textsf{sledgehammer}
tool~\cite{Blanchette2011}, and counterexample generators like
\textsf{nitpick}. We can directly harness many of these proof
facilities by our transfer based proof tactics. Finally, Isabelle has
been integrated into Eclipse in the form of
\emph{Isabelle/Eclipse}\footnote{\url{http://andriusvelykis.github.io/isabelle-eclipse/}},
an IDE which we reuse in Symphony.

%We chose Isabelle for several reasons. It is \emph{principled}, being
%based on the \emph{LCF} architecture, where proofs are
%correct-by-construction with respect to a secure logical core. It is
%also mature, consisting of a large library of mathematical structures,
%and integrating a number of powerful proof facilities, including
%support for \emph{Z3} and automated theorem provers through the
%\emph{Sledgehammer} tool. It is also based on \emph{Higher Order
%  Logic}, which makes it ideal for embedding a language. Moreover
%Isabelle also has been integrated into Eclipse in the
%\emph{Isabelle/Eclipse}~\footnote{\url{http://andriusvelykis.github.io/isabelle-eclipse/}}
%environment, which we make use in Symphony.

%% file: sections/tppog.tex
\section{PO Generation and Discharge in Symphony}
\label{sec:pog_tp}

%The \ac{pog} and \ac{tpp} combine to verify properties of a given \ac{cml}
%model by employing a two-step approach. The \ac{pog} examines the model and
%generates proof goals (the \acp{po}) that, when discharged, ensure the
%correctness of the model. The \ac{tpp}~\cite{COMPASSD33.2b} is responsible for
%discharging these goals and does so by interacting with the Isabelle
%theorem prover. 

The Symphony \ac{pog} is an extension of the Overture \ac{pog} for
the \ac{vdm}~\cite{couto&13a}, and therefore many proof goals generated
are derived from \ac{vdm}. However, the \ac{pog} has been developed
with an extensible visitor~\cite{DESIGNPAT95} based architecture that
will enable the addition of further goals as they are researched. The
current proof goals fall broadly into the following categories: safe
usage of partial operators; safe usage of functions with
pre-conditions; type compatibility due to union types, type invariants
and subtypes; and satisfiability of implicitly defined functions and
operations.

To illustrate the use of \acp{po} in Symphony, we present a simple
example based on a well known partial operator case: division by
zero. Consider the following \ac{cml} function:

\begin{cml}
division : int * int -> real
division (x,y) == x / y
\end{cml}
\vspace{-3ex}

\noindent For this function, the \ac{pog} generates an obligation
stating that, for all inputs to the function, the value of the
divisor (\texttt{y}) will not be 0, thus ensuring the function executes
successfully.  This is represented in the logical formula below:

\vspace{-1ex}
\begin{cml}
PO1: forall x:int, y:int & (y <> 0) 
\end{cml}

\vspace{-3ex}

\begin{figure}
    \centering
    \includegraphics[width=0.48\textwidth]{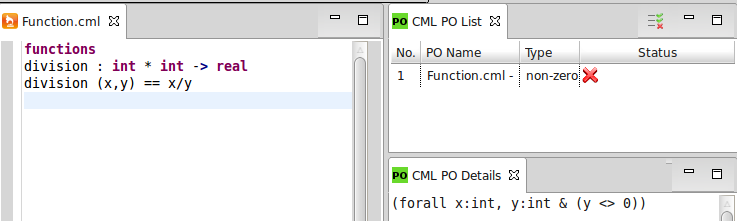}
    \caption{Failed PO discharge.}
    \label{fig:discharge-fail}
    \vspace{-4ex}
\end{figure}

\noindent This states that for all variables \texttt{x} and \texttt{y}
of type \texttt{int}, \texttt{y} is not equal to 0. This is not
satisfiable, and so an attempt to discharge this \ac{po} will fail as
shown in \autoref{fig:discharge-fail}. The function must be enriched
with a pre-condition, using the \textbf{\texttt{pre}} keyword, in order for the
discharge to be possible:

\vspace{-1ex}
\begin{cml}
division : int * int -> real
division (x,y) == x / y
pre y <> 0
\end{cml}

\vspace{-3ex}

\begin{figure}[h]
    \centering
    \includegraphics[width=0.48\textwidth]{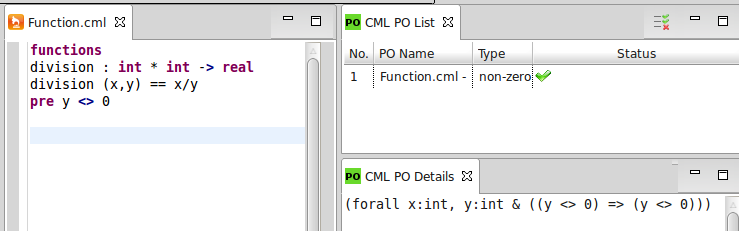}
    \caption{Successful PO discharge.}
    \label{fig:discharge-ok}
    \vspace{-2ex}
\end{figure}

\noindent The additional information offered by the pre-condition alters the
\ac{po} and now the theorem prover plug-in is able to  discharge the revised
\ac{po} as shown in \autoref{fig:discharge-ok}. It is of course not possible to
prove that an arbitrary integer is different from zero, but it is trivial
to prove that a non-zero integer is different from zero.

\noindent The aforementioned example was trivial but, in general, it is quite
important to ensure that, when adding a pre-condition, said pre-condition is
sufficient to allow the discharge of any \acp{po} generated for the function.

The addition of the pre-condition has another effect. One must now ensure
that, whenever the function is called its pre-condition is respected. Therefore, a new
kind of \ac{po} is generated. Consider the following function and PO: 

\vspace{-1ex}
\noindent\begin{minipage}{0.47\textwidth}
\begin{cml}
divby2: int -> real
divby2 (x) == division(x,2)
\end{cml}
\vspace{-2ex}
\end{minipage}

\begin{cml}
PO2: forall x : int & pre_division(x,2)
\end{cml}
\vspace{-2ex}

\noindent Since \texttt{divby2} calls \texttt{division}, a \ac{po}
is generated to ensure that the pre-condition of \texttt{division},
given by \texttt{pre\_division}, is satisfied. This kind of obligation,
called \texttt{pre-condition obligation} is generated at all points in the model
where the function is called.

While the example shown was very simple, pre-conditions (and invariants)
can be as complex as necessary. They are expressed in the functional subset
\ac{cml} and thus have the full expressive power of CML's first-order
logic. In addition to helping ensure consistency of the model, pre-conditions and 
invariants are also used to specify additional properties 

In fact, the methodology we propose is based precisely on specifying desired
properties and requirements of a model through pre- and post-conditions as
well as invariants. The \acp{po}, once generated and discharged, stand as
proof that the model respects the specified properties.

Discharging \acp{po} is the task of the \ac{tpp}. At its core, the \ac{tpp}
consists of a mechanised semantic model for \ac{cml} within \isabelleutp. It is
essentially a deep embedding of \ac{cml}, in that we give an explicit semantics
to each of the operators of \ac{cml} processes within Isabelle.  

\begin{figure}
    \centering
    \includegraphics[width=0.45\textwidth]{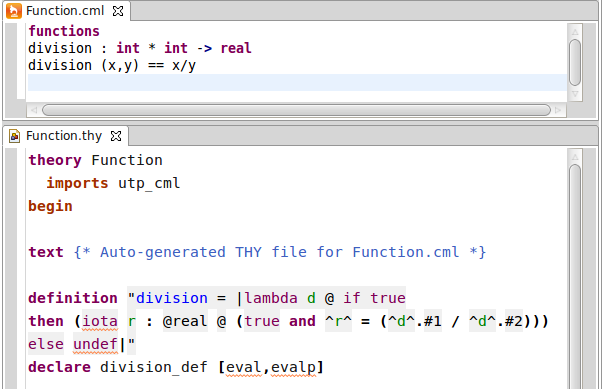}
    \label{fig:pog}
    \caption{Auto-generated theory files.}
    \label{fig:cml-isa}
    \vspace{-3ex}
\end{figure}

The \ac{tpp} will process a \ac{cml} model and its associated \acp{po} and
automatically generate Isabelle theory files for them (see
\autoref{fig:cml-isa}). These theory files can then be submitted
to Isabelle for discharging through various automated proof tactics such as
\textsf{auto} and \textsf{sledgehammer}, or the \textsf{cml\_tac} tactic that
maps a \ac{cml} formula onto a HOL formula.

Because the \ac{tpp} connects to the \emph{Isabelle/Eclipse} plug-in, the full
functionality of that plug-in and, by extension, Isabelle is available to the
user . This includes the ability to write and discharge model-specific
conjectures directly in the Isabelle encoding of the model.  However, to
perform this kind of work requires significant knowledge of Isabelle and its
syntax.

Therefore, the \ac{pog} will be the primary source of goals to discharge.
Furthermore, the \ac{tpp} offers a fully automated mode of interaction with
Isabelle where users simply choose which \ac{po} to discharge and all inner
workings (such as tactic selection and result collection) are hidden from them.

We envisage two main functionalities for the plug-ins \emph{Quick check} and \emph{Proof Session}. \emph{Quick check} will be a fully automated process, simply 
presenting a list of \ac{pos} (including their predicates) in \ac{cml} and 
linked to the relevant model elements.  The process of generating \acp{po} is 
quick, therefore this may be performed frequently during initial model development,
to gain useful feedback about the model.
%         In either case, the purpose of the quick
%         check phase is to provide a fast, fully automated analysis that can
%         help guide the user during development towards a correct model. Because
%         of this, all results from the quick check phase are discarded.

The \emph{Proof Session} will be the main functionality of the plug-ins. It
creates a snapshot of the model (a timestamped,
read-only copy of the model's \ac{cml} sources), generates \acp{po} and
translates the model and \acp{po} into Isabelle theories.  The \acp{po} are
displayed in a similar manner to the quick check version but can now be
submitted to the \ac{tpp} for discharge. At the moment only
\textsf{cml\_tac} is available, though we hope to enable automated use of
additional tactics when attempting to discharge \acp{po}.  Regardless, the
output of a proof attempt will be captured from Isabelle and displayed to the
user. Also, the results of a proof session will be stored along with the
model snapshot, thus verifying the model's correctness.

These two functionalities combine to form the following work-flow: as
a user works on a model, he can quick check for \acp{po} as a way to
gain early insights into the of the model. Each \ac{po} can be seen of
as a possible inconsistency and merely by manual inspection they can
guide the user in terms of adding necessary pre-conditions or guards
to the model.

Once a set of changes has been completed, the user may use the proof session
functionality to verify the model's correctness.  Each set of \acp{po} and
their associated proofs are only valid for the particular version of the model
they were generated from so it makes little sense to attempt manual proofs on
a volatile model. Regardless of when it is attempted, the proof session for the
average user will be fully automated. The user simply initiates a proof session
and selects \acp{po} for discharging either manually or in batch. Typically
some \acp{po} will be successfully discharged whereas others will fail to
discharge.  These should indicate a problem with the model and action must be
taken by the user (for example, by adding a guard or correcting program logic)
to alter the model in a way that allows the \ac{po} to be discharged. Then, the
set of completed \ac{po} goals can be used as a formal proof of the constituent
system's correctness.

For advanced users who are comfortable interacting with \emph{Isabelle/Eclipse}
directly,  the full theorem proving perspective gives them direct access to the
tool so that manual proofs may be attempted. Users can also  specify and
discharge additional model-specific conjectures.

%% file: sections/example.tex
\section{Verification of Example Constituent System}
\label{sec:example}

We illustrate the use of the \toolpl\ \ac{pog} and \ac{tpp} with a simple example CS from a Railway Signal \ac{sos}. The \ac{sos} in question aims to ensure the safe and correct movement of trains on a section of railway tack. Naturally such a \ac{sos} poses several dependability concerns and the integrator of the \ac{sos}  requires several safety properties to hold throughout the life of each of the systems.

%In the SoS, we consider several types of signal indicators -- each potentially owned and manufactured by different companies.  
%The \ac{sos} architecture structure is defined in the SysML block definition diagram in Figure~\ref{fig:dwarf-bdd}. 
The \emph{Railway Signal} \ac{sos} comprises several constituents including a \emph{Route Rule Engine}, several \emph{Track Actuators}, \emph{Trains} and \emph{Dwarf Signal} systems. In this paper, we look at one of the constituent systems -- the \emph{Dwarf Signal} system -- in detail and consider the safety properties of that system.

\begin{figure}
	\centering
	\includegraphics[width=0.35\textwidth]{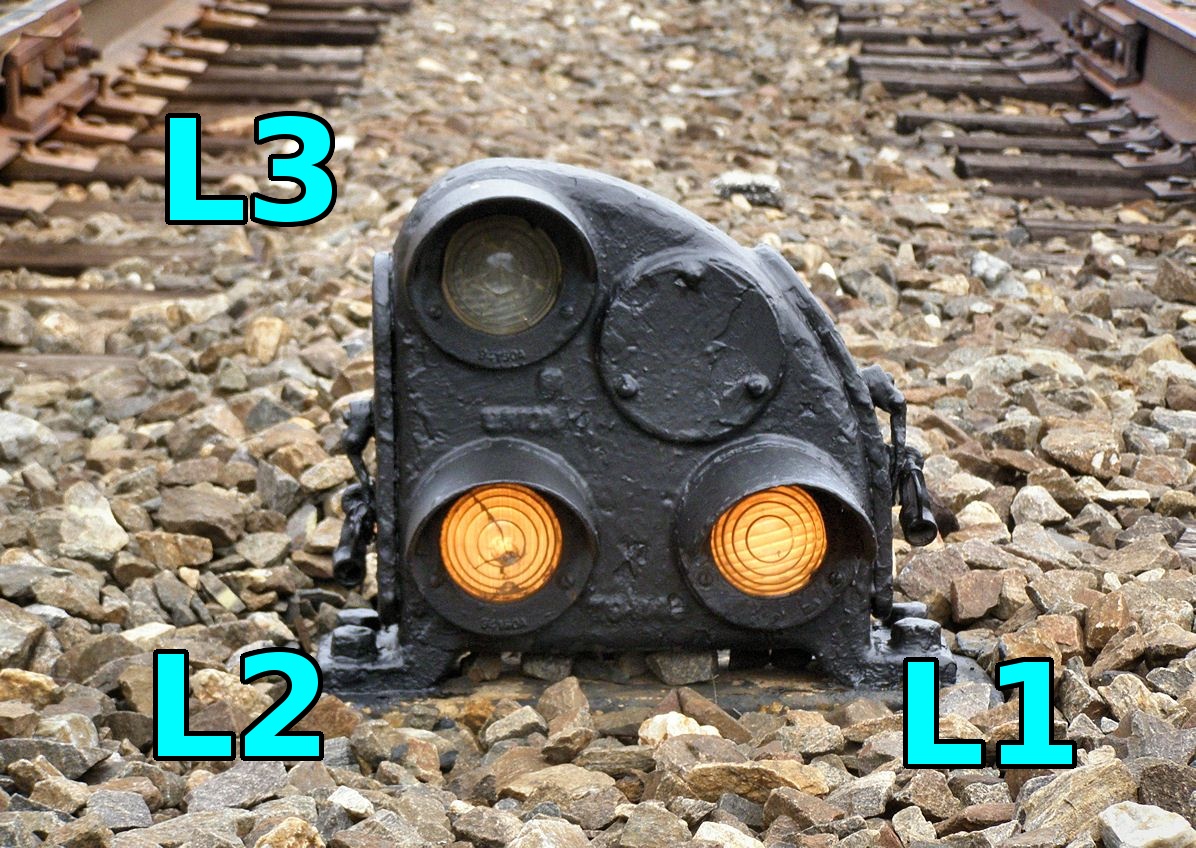}
	\caption{Picture of railway signal, with lamps indicated}
	\label{fig:dwarf-pic}
        \vspace{-4ex}
\end{figure}

From the perspective of the SoS integrator, there is a requirement that the procured constituent systems provide a safe service. The constituent system designer must, therefore, provide evidence of this safety. Using model-based techniques, we define a formal model of the \emph{Dwarf Signal} -- which may be used as a contract to which the the signals must conform. The \emph{Dwarf} model used in this paper is based upon that introduced in~\cite{Foster13}, and a typical signal may be seen in Figure~\ref{fig:dwarf-pic}.

%\begin{figure}[h!]
%	\centering
%	\includegraphics[width=0.48\textwidth]{figures/railway_bdd}
%	\caption{SysML block definition diagram for case study structure}
%	\label{fig:dwarf-bdd}
%\end{figure}
%\vspace{-0.7cm}

The \emph{Dwarf Signal} model is defined in CML with several datatypes, functions, a single \texttt{Dwarf} process with state variables, operations and actions. The main datatype, \texttt{DwarfType} shown below, has several fields relating to the transitions which are to be made in the \emph{Dwarf Signal}. For example, the \texttt{currentstate} field dictates the collection of lamps currently lit, and the \texttt{desiredproperstate} field represents the next state the \emph{Dwarf Signal} should reach. The set of possible signal states that may be reached is defined by the \texttt{ProperState} datatype, which is constrained to be one of for constant values: \texttt{dark, stop, warning} and \texttt{drive} -- each a set of lamps.

\vspace{-2ex}
\noindent\begin{minipage}{0.47\textwidth}
\begin{cml}
types
 LampId      = <L1> | <L2> | <L3>
 Signal      = set of LampId
 ProperState = Signal
 inv ps == ps in set {dark, stop, warning, drive}

 DwarfType :: lastproperstate    : ProperState
              turnoff            : set of LampId 
              turnon             : set of LampId
              laststate          : Signal
              currentstate       : Signal
              desiredproperstate : ProperState
 inv d == NeverShowAll(d) and MaxOneLampChange(d) and ForbidStopToDrive(d) and DarkOnlyToStop(d)and DarkOnlyFromStop(d)
		   
values
 dark: Signal    = {}
 stop: Signal    = {<L1>, <L2>}
 warning: Signal = {<L1>, <L3>}
 drive: Signal   = {<L2>, <L3>}
\end{cml}
\end{minipage}
\vspace{-3ex}

\noindent There are several safety properties to which the \emph{Dwarf Signal} must adhere. These are defined in terms of functions referred to in the \texttt{DwarfType} invariant -- including, for example, \texttt{NeverShowAll} which requires that the \texttt{currentstate} should never have all three lamps lit. The \texttt{Dwarf} process, outlined below has a single state variable: \texttt{dw} of type \texttt{DwarfType}, and four operations: \texttt{Init}, which initialises the \texttt{dw} state variable; \texttt{SetNewProperState}, allowing the next desired \texttt{properstate} to be set; and two operations for changing the lamps lit in the signal -- \texttt{TurnOn} and \texttt{TurnOff}.

\vspace{-1ex}
\noindent\begin{minipage}{0.47\textwidth}
\begin{cml}
process Dwarf = begin 
state 
  dw : DwarfType

operations
  Init : () ==> ()
  Init() == (...)

  SetNewProperState: (ProperState) ==> ()
  SetNewProperState(st) == (...)

  TurnOn: (LampId) ==> ()
  TurnOn(l) == (...)

  TurnOff : (LampId) ==> ()
  TurnOff(l) == (...)
 ... end
\end{cml}
\end{minipage}
\vspace{-3ex}

\noindent Each operation is defined in more detail in terms of pre- and post-conditions, dictating the conditions in which the operation may be called and the guarantees it makes if those conditions are met. The \texttt{Init} operation, defined in more detail below, has a body which initialises the \texttt{dw} state variable, with a post-condition requiring that various fields of the \texttt{dw} variable are updated. The operation body -- an assignment to the \texttt{dw} state variable -- must respect the safety properties of the \emph{Dwarf Signal}, in the form of the type invariant described above.

% In the $SetNewProperState$ operation, defined in more detail below, the pre-condition requires that before the operation is evaluated, the Dwarf $currentstate$ is the $desiredproperstate$ (such that the Dwarf has made the previous state change) and that the new state provided as a parameter $st$ is different to the $currentstate$. The post-condition for the operation requires that various fields of the $dw$ variable are updated.

\vspace{-1ex}
\begin{cml}
Init : () ==> ()
Init() ==
  dw := mk_DwarfType(stop, {}, {}, stop, stop, stop) 
post dw.lastproperstate = stop and dw.turnoff = {} 
       and dw.turnon = {} and dw.laststate = stop 
       and dw.currentstate = stop 
       and dw.desiredproperstate = stop
\end{cml}
\vspace{-2ex}

%SetNewProperState: (ProperState) ==> ()
%SetNewProperState(st) ==
%  dw := mk_DwarfType(dw.currentstate,
%  	                 dw.currentstate \ st,
%  	                 st \ dw.currentstate,
%  	                 dw.laststate,
%  	                 dw.currentstate,
%  	                 st) 
%pre dw.currentstate = dw.desiredproperstate and
%    st <> dw.currentstate    
%post dw.lastproperstate = dw~.currentstate and
%     dw.turnoff = dw~.currentstate \ st and
%     dw.turnon  = st \ dw~.currentstate and
%     dw.laststate = dw~.laststate and
%     dw.currentstate = dw~.currentstate and
%     dw.desiredproperstate = st

\noindent The remainder of the CML operations are defined in a similar manner -- with pre- and post- conditions. In addition to these operation definitions, the CML model contains \textit{actions} which dictate the ordering of internal events and operation calls. At present, the \ac{pog} does not handle these features of CML, and thus they are omitted from this paper.

Executing the \tool\ \ac{pog}, we obtain several \acp{po}, which are generated by the \texttt{Init}, \texttt{SetNewProperState}, \texttt{TurnOn} and \texttt{TurnOff} operations. The \acp{po} fall into two \ac{po} types: ensuring that the postcondition holds given the body of the operation; and ensuring subtype consistency. It is the second of these which ensures that the \texttt{DwarfType} type invariant (and thus the safety properties of the \emph{Dwarf Signal}) holds when setting a new value of the \texttt{dw} variable. The generated subtype \acp{po} (\texttt{PO1} and \texttt{PO2}) and postcondition \ac{po} (\texttt{PO3}) for the \texttt{Init} operation are shown below. 

\vspace{-1ex}
\begin{cml}
PO1: inv_ProperState(stop) 

PO2: ((inv_DwarfType(mk_DwarfType(stop, {}, {}, stop, stop, stop)) and inv_ProperState(stop)) and inv_ProperState(stop))

PO3: (((dw.lastproperstate) = stop) and (((dw.turnoff) = {}) and (((dw.turnon) = {}) and (((dw.laststate) = stop) and (((dw.currentstate) = stop) and ((dw.desiredproperstate) = stop))))))
\end{cml}
\vspace{-2ex}

Using the \toolpl, we generate these \acp{po}, and attempt to discharge them. In Figure~\ref{fig:dwarf-pos} below, we show \tool\ in the \ac{pog} perspective with the \acp{po} represented in the Isabelle syntax used by the \ac{tpp}. In the figure, the list of \acp{po} is given in the right hand pane, with a pane showing the \ac{po} definition in CML below. In the figure, we see that several of the \acp{po} have been discharged -- these relate to the $Init$ operation above -- as denoted by the green ticks.

\begin{figure}[h]
	\centering
	\includegraphics[width=0.48\textwidth]{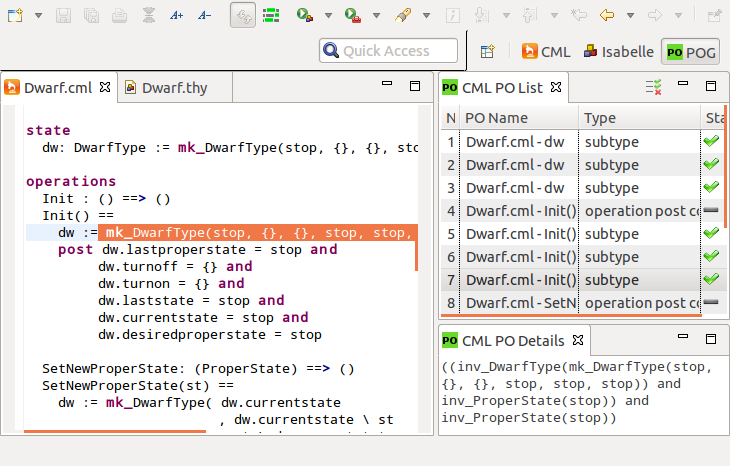}
	\caption{Progress on POs generated for Dwarf model}
	\label{fig:dwarf-pos}
    \vspace{-0.6cm}
\end{figure}

By discharging all \acp{po} for the \emph{Dwarf Signal} model, we provide a contractual model which is verified to be both internally consistent and, through encoding the safety properties which must be met by a signal, is safe with respect to the requirements placed on that contract. At present, whilst those \acp{po} shown are successfully discharged by the \tool\ \ac{tpp}, several are not. These relate to those \acp{po} which rely upon the value of the \emph{Dwarf} process state variable \emph{dw} at a given point of time. This may be either an issue with the \ac{po} expressions themselves (where the VDM-based \ac{po} expressions require further adaption to CML), or due to the early stage of development of the \ac{tpp} proof tactics. We discuss these areas as future work in the next section.

%\begin{itemize}
%    \item What this means for the example
%%    \item Some PO fails to discharge
%%    \item Change model
%%    \item Discharge again
%%    \item Success!
%\end{itemize}

%% file: sections/conclusion.tex
\section{Conclusion and Future Work}
\label{sec:conc}

In this paper, we have outlined two \toolpl\ plugins which enable
automated proof support for \ac{cml} \acp{po}. The \tool\ \ac{pog}
reuses and expands upon the Overture \ac{pog}, also resulting in
improvements in the Overture Tool. The \ac{tpp} is the first attempt
at tightly integrating a theorem prover into the VDM family of tools,
providing a more useful tool for users wishing to discharge \ac{cml}
\ac{pos} and general theorems. We have also shown how both plug-ins
can be used in combination to formally verify the integrity of
constituent systems of an \ac{sos} specified with \ac{cml}. There are clearly many areas of future work, both short-term improvements to the two plugins,  and also longer-term directions and scoping of the work in the fields of \ac{sos} and dependability-related issues. Below, we discuss several such directions for further work.

This paper demonstrated the verification of constituent system
\emph{models}. An interesting issue would be the verification of an
actual implementation, with respect to realistic system
properties. Whilst clearly not in the scope of this paper, we would
consider the work of this paper in context with other system
engineering activities. In particular; positioning constituent system
verification with respect to the work of Holt et al~\cite{Holt12} on
SoS requirements engineering and the specification of SysML contracts
and translation to CML~\cite{Bryans2014a} may provide the means to
more realistic property verification.

The current CML \ac{tpp} focuses on VDM-style proof obligations
which deal with issues such as subtyping and internal consistency. In
the future we will extend this with a more comprehensive calculus,
such as a Hoare logic~\cite{COMPASSD23.4d} or a weakest precondition
calculus, which would both expand on the existing proof
obligations. This would also allow us to reason directly about CML
process and state behaviour, and therefore provide fuller support for
reasoning about contracts.

Another extension to the proof obligations relates to scaling our approach
from the level of constituent systems to the \ac{sos}-level, thus ensuring that
the verified constituent systems  interact in a manner that ensures the desired
behaviour of the overall \ac{sos}.

Just as pre-conditions and invariants can be used to specify the
properties that the \ac{pog} and \ac{tpp} verify, we need a mechanism
that allows these tools to reason about correctness at the \ac{sos}
level. We see two distinct possibilities here: the first is to
introduce a new \ac{cml} construct that allows one to specify
invariants over the entire \ac{sos}, thus being able to ``see'' inside
all constituents. The second approach is to take the existing \acp{po}
that verifies a system and use them to also verify the interface of a
constituent. Afterwards, one must establish a means by which these
verified interfaces can be combined to establish global \ac{sos}
properties.  Of the two approaches, the second one seems closer to the
spirit of \ac{sos} engineering, and we believe CS refinement provides
a way forward here.

We are also currently working on a tool for CS refinement, which
combines with the theorem prover and can be used to formally
demonstrate contractual satisfaction. This will reuse the \ac{pog} to
enumerate and discharge refinement provisos which must often be
satisfied to ensure validity of a refinement step. Refinement will be
principally supported by Isabelle, though we are also exploring the
use of model generation tools to aid automation. For example, the
Maude rewriting logic engine~\cite{Clavel2002187} has previously been
applied to automated refinement~\cite{griesmayer13refine}, which we
hope to adapt for CML. Such advances over the current technology are
feasible because of our extensible approach to semantics provided by
UTP.

Finally, though both plug-ins presented here are still at an early development
stage, work is ongoing on various improvements. While the \ac{tpp} and its
associated Isabelle theories support a significant subset of \ac{cml} (types,
expressions, functions, and operations), work is ongoing on increasing the
coverage of the plug-in. The proof tactics are also under further development
in order to discharge increasingly complex goals. Moreover we wish to expose
more of Isabelle's native proof facilities in the \ac{tpp}, such as
\textsf{sledgehammer} and \textsf{nitpick}, so as to bring their full weight to
bear in discharging or refuting proof obligations. Parallel to this there is
work to formalise the proof obligations in Isabelle with respect to the
\ac{cml} semantics. Finally, we hope to produce guidance to the user of how to
interpret failure when a PO cannot be discharged.

%In general, though, we believe the tool is at a stage where it can be used 
%successfully to validate supported specifications.